\documentclass[1p]{elsarticle}
\usepackage[utf8]{inputenc}
\usepackage[english]{babel}
\usepackage[T1]{fontenc}
\usepackage{amsmath}
\usepackage{graphicx}
\usepackage{url}
\usepackage{units}

\begin{document}
\biboptions{compress}
\title{Investigation of $\alpha$-induced reactions on the $p$ nucleus $^{168}$Yb}

\author[col]{L.~Netterdon\corref{cor1}}
\ead{lnetterdon@ikp.uni-koeln.de}
\cortext[cor1]{Corresponding author}
\author[dem]{P.~Demetriou\fnref{fn1}}
\author[col]{J.~Endres}
\author[ptb]{U.~Giesen}
\author[ato]{G.~G.~Kiss}
\author[col]{A.~Sauerwein}
\author[ato]{\\T.~Sz\"ucs}
\author[col]{K.~O.~Zell}
\author[col]{A.~Zilges}

\address[col]{Institut f\"ur Kernphysik, Universit\"at zu K\"oln, Z\"ulpicher Stra{\ss}e 77, D-50937 K\"oln, Germany}
\address[dem]{Institute of Nuclear Physics, NCSR ``Demokritos'', GR-153.10 Aghia Paraskevi, Athens, Greece}
\address[ptb]{Physikalisch-Technische Bundesanstalt (PTB), Bundesallee 100, D-38116 Braunschweig, Germany}
\address[ato]{Institute for Nuclear Research (ATOMKI), H-4001 Debrecen, Hungary}

\fntext[fn1]{Present address: NAPC-Nuclear Data Section, International Atomic Energy Agency, A-1400 Vienna, Austria}

\begin{abstract}

Cross sections for the $^{168}$Yb($\alpha$,$\gamma$)$^{172}$Hf and $^{168}$Yb($\alpha$,n)$^{171}$Hf reactions were measured by means of the activation method using $\alpha$ particles with energies between \unit[12.9]{MeV} and \unit[15.1]{MeV}. The spectroscopy of the $\gamma$ rays emitted by the reaction products was performed using three different HPGe detector types, namely clover-type high-purity germanium detectors, a low-energy photon spectrometer detector, and a coaxial high-purity germanium detector. The results were compared to Hauser-Feshbach statistical model calculations. Within certain assumptions, astrophysical conclusions could be drawn concerning the production of the $p$ nucleus $^{168}$Yb. The data in this work can serve as a contribution to the current very fragmentary experimental data base for charged-particle induced reactions. In addition, the absolute intensity for nine $\gamma$-ray transitions following the electron capture decay of $^{171}$Hf could be derived.

\end{abstract}

\begin{keyword}
nuclear astrophysics \sep $p$ process \sep $\alpha$-induced reactions \sep $^{168}$Yb \sep measured cross section \sep activation method  
\end{keyword}

\maketitle
\section{Motivation}
\label{sec:motivation}
Almost all stable nuclei heavier than iron are synthesized by neutron capture reactions during the $s$ and $r$ process \cite{B2FH,Kaeppeler11,Arnould07}. However, about 35 neutron-deficient nuclei between $^{74}$Se and $^{196}$Hg can be produced neither by the $s$ nor the $r$ process. They are referred to as $p$ nuclei \cite{Woosley78,Arnould03}.

According to current knowledge, different mechanisms contribute to the production of $p$ nuclei. Several processes are suggested such as the $\gamma$ process \cite{Woosley78,Rayet95}, the $rp$ process \cite{Schatz98}, the $\nu p$ process \cite{Froehlich06}, and the $pn$ process \cite{Goriely02}. The processes are subsumed in here under the heading $p$ process. The current understanding is that the majority of the $p$ nuclei are produced by photodisintegration reactions within the $\gamma$ process occurring on existing $s$- and $r$- process seeds \cite{Rayet95}. It was found that these $\gamma$-induced reactions may occur within O/Ne burning layers of core-collapse supernovae at temperatures of about \unit[2]{GK} to \unit[3]{GK} \cite{Woosley78, Rayet95, Arnould76}. The supernova shock wave allows photodisintegration reactions and subsequent $\beta$ decays to produce $p$ nuclei. Recent calculations have shown that a significant amount of $p$ nuclei can also be produced during type Ia supernovae \cite{Travaglio11}. Furthermore, it was shown that lighter $p$ isotopes up to $^{96,98}$Ru can be co-produced in a high-entropy wind scenario during type II supernovae \cite{Farouqi09}.

Since experimental data are scarce, calculations regarding the extensive $\gamma$-process reaction network rely almost completely on theoretical reaction rates predicted by the Hauser-Feshbach (HF) statistical model \cite{Hauser52}. Although the HF model itself is well-established, major uncertainties stem from the nuclear physics input parameters entering these calculations. Important experimental efforts have been made during the last years, especially for the intermediate and heavy mass region using the activation method \cite{Yalcin09,Gyuerky10,Kiss11,Filipescu11,Sauerwein11,Dillmann11,Halasz12}, which is also used in the present work, as well as the in-beam technique with high-purity germanium (HPGe) detectors \cite{Harissopulos01,Galanopoulos03,Sauerwein12} and the 4$\pi$ summing method \cite{Tsagari04, Spyrou07}. However, especially the description of the $\alpha$+nucleus optical model potential (OMP) remains a problem, experimental data of $\alpha$-capture reactions on heavy nuclei for low energies are often overestimated by theoretical predictions. The $\alpha$-OMP determines deflections and branchings in the $\gamma$-process path and thus has a direct impact on the $p$-nuclei abundances. Especially the mass range $150~\leq~A~\leq165$ remains problematic to be reproduced. Moreover, it was pointed out in Refs.~\cite{Rauscher06,Rapp06} that certain reaction rates have a direct influence on the final abundance of the $p$ nuclei. One of the proposed reactions to be studied experimentally in Ref.~\cite{Rauscher06}, $^{168}$Yb($\alpha$,$\gamma$)$^{172}$Hf, is subject of the present study.
Since $^{168}$Yb is a $p$ nucleus and close the mass range $150~\leq~A~\leq165$ which remains problematic to be reproduced, low-energy $\alpha$-induced data on this nucleus might serve as an important contribution to improve the situation regarding the $\alpha$-OMP.  

\begin{figure}[thb]
\centering
 \includegraphics[width=\textwidth]{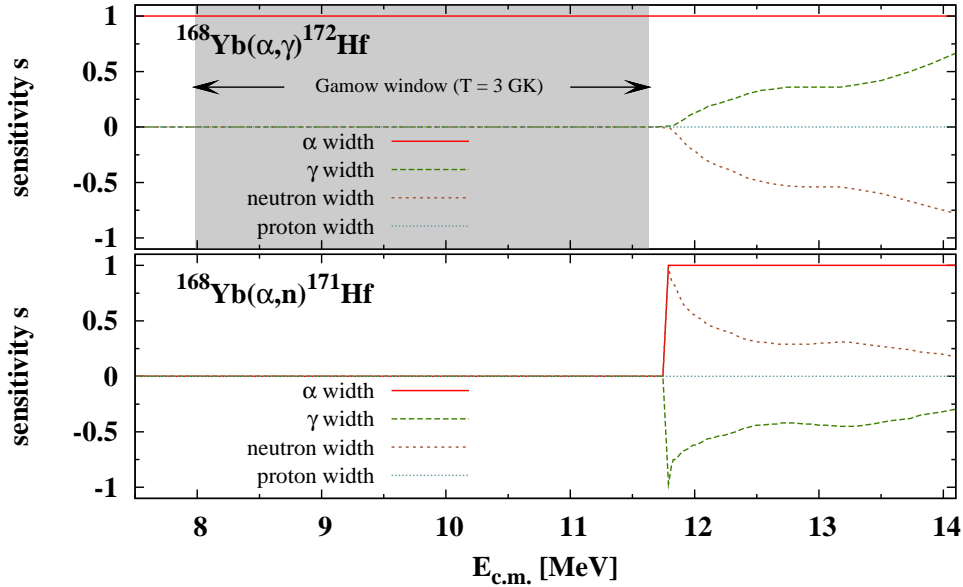}
\caption{(Color online) Sensitivity of laboratory cross sections of the $^{168}$Yb($\alpha$,$\gamma$) (upper panel) and $^{168}$Yb($\alpha$,n) (lower panel) reaction to variations of various widths as a function of center-of-mass energies \cite{Rauscher12}. The widths were varied by a factor of two. Within the Gamow window, the ($\alpha$,$\gamma$) cross section is only sensitive to variations of the $\alpha$ width. After the neutron emission channel opens, \textit{i.e.} within the measured energy range, the laboratory cross sections of both reactions show a non-negligible sensitivity to the $\gamma$ width and the neutron width.}
\label{fig:sensitivity}
\end{figure}

Within the astrophysically relevant energy range between \unit[8]{MeV} and \unit[11.6]{MeV} for a temperature of \unit[3]{GK} \cite{Rauscher10}, the cross section of the $^{168}$Yb($\alpha$,$\gamma$) reaction is only sensitive to the $\alpha$ width \cite{Rauscher12}, see Fig.~\ref{fig:sensitivity} (upper panel). This means, that only changes of the $\alpha$ width and, thus, the $\alpha$-OMP result in changes of the cross section. A variation of the other nuclear properties leaves the cross section unchanged. A negative value of a sensitivity implies, that the cross section varies inversely proportional to the respective width. However, the ($\alpha$,$\gamma$) cross sections within the Gamow window are too small to be measured using the activation method within a reasonable time. Hence, the measurements had to be performed at energies above the Gamow window. Within the experimentally accessible energy range above the neutron emission threshold, the cross section is also sensitive to the neutron and $\gamma$ width. This also holds for the ($\alpha$,n) reaction, which was measured simultaneously. Although the sensitivity of the ($\alpha$,n) reaction to the input parameters beside the $\alpha$-OMP is much less compared to the ($\alpha$,$\gamma$) case, it cannot be ignored especially in the lower energy region, see Fig.~\ref{fig:sensitivity} (lower panel).

In this work, the ($\alpha$,n) and ($\alpha$,$\gamma$) reactions are investigated at center-of-mass energies between $E_{\mathrm{c.m.}}$~=~\unit[12.5]{MeV} and \unit[14.7]{MeV} using the activation method. The measured energy range is located above the Gamow window for the ($\alpha$,$\gamma$) reaction. The irradiation of the targets was performed using the cyclotron of the Physikalisch-Technische Bundesanstalt (PTB) in Braunschweig, Germany \cite{PTB}. 

In Sec.~\ref{sec:experiment} the experimental procedure is presented, followed by the data analysis which is explained in Sec.~\ref{sec:dataanalysis}. The experimental results and their discussion are given in Sec.~\ref{sec:results}.

\section{Experiment}
\label{sec:experiment}

\subsection{Investigated reactions}
\label{subsec:reactions}

\begin{figure}[thb]
\centering
 \includegraphics[width=0.8\textwidth]{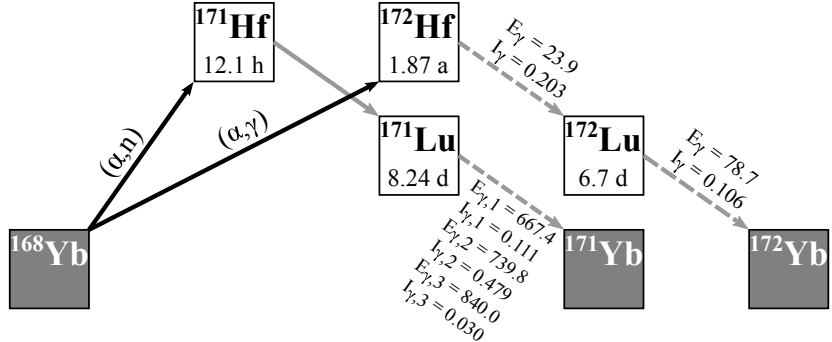}
\caption{Illustration to show the measured reactions and their decay products. Stable nuclei are illustrated as gray boxes, whereas unstable nuclei are illustrated as white boxes. Only the nuclei relevant for the two reactions are shown here. The investigated reactions are depicted by the thick arrows while the decays used for data analysis are indicated by dashed arrows. The $\gamma$-ray energies are given in keV. Note, that for better readability, the errors of the $\gamma$-ray intensities and half-lives are not shown here, see Table~\ref{tab:decaydata} for more information. Half-lives, $\gamma$-ray energies, and $\gamma$-ray intensities are taken from Ref.~\cite{NNDC}.}
\label{fig:reaction}
\end{figure}

In total, six targets were irradiated using incident $\alpha$-particle energies between \unit[12.9]{MeV} and \unit[15.1]{MeV}. The beam current was limited to $\approx$ \unit[600]{nA} to guarantee thermodynamical target stability. The activation runs lasted between 5 hours for the higher beam energies and 20 hours for the lowest beam energy. The Q value of the ($\alpha$,n) reaction is \unit[($-11796.5~\pm~29.0$)]{keV}, whereas the Q value for the ($\alpha$,$\gamma$) reaction amounts to \unit[($-2754.3~\pm~24.5$)]{keV} \cite{QCalc}.
 Figure \ref{fig:reaction} gives an overview of the investigated reactions in this experiment and their decay products. The $\gamma$-ray transitions following the electron-capture decay of $^{171}$Hf could not be used to determine the cross section of the ($\alpha$,n) reaction. Their absolute $\gamma$-ray intensities are unknown. However, the normalization factor to calculate these $\gamma$-ray intensities could be derived from this experiment, see Sec.~\ref{sec:intensity}. For the determination of the ($\alpha$,n) cross section, the decay of $^{171}$Lu was used. The cross section of the ($\alpha$,$\gamma$) reaction was measured using the decays of $^{172}$Hf and $^{172}$Lu. The decays used in this experiment are indicated by dashed arrows in Fig.~\ref{fig:reaction}.

\subsection{Target characterization}
\label{subsec:target}

\begin{figure}[tb]
\centering
\includegraphics[width=\textwidth]{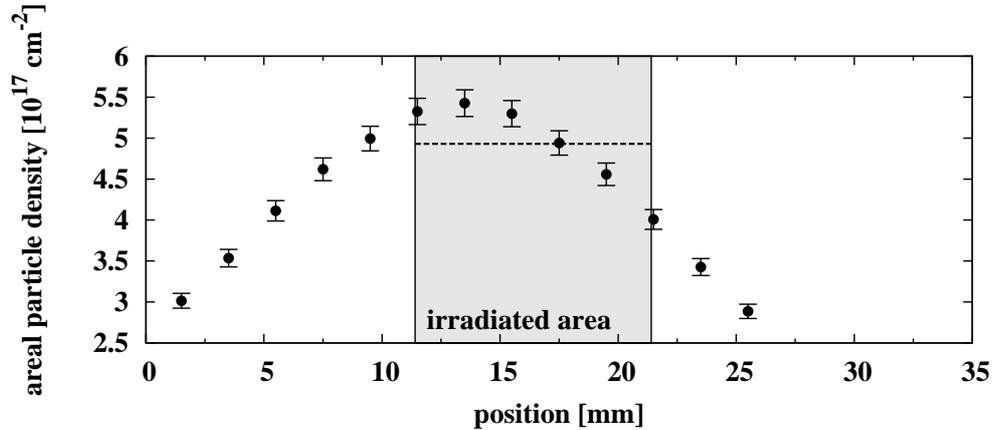}
\caption{Areal particle density of $^{168}$Yb target nuclei measured every \unit[2]{mm} over the whole target using the RBS method. The irradiated area is marked in light gray where the target thickness varies up to \unit[25]{\%}. Over this area the weighted average was calculated, depicted by a dashed line, and used later on for data analysis under the assumption of a uniform irradiation.}
\label{fig:rbs}
\end{figure}

The targets were prepared using Yb$_\mathrm{2}$O$_\mathrm{3}$ which was reduced by Hf and subsequently evaporated onto \unit[1]{mm} thick high-purity Al backings. The isotopic enrichment in $^{168}$Yb amounts to \unit[(35.2~$\pm$~0.5)]{\%} (for one target \unit[(13.7~$\pm$~0.5)]{\%}). Target thicknesses were measured using the Rutherford Backscattering Spectrometry (RBS) facility at the RUBION Dynamitron-tandem accelerator at the Ruhr-Universit\"at Bochum. For this measurement, the targets were irradiated with $^4$He$^+$ ions which had an energy of \unit[2]{MeV}~$\pm$~\unit[1]{keV} with a current of about \unit[15]{nA}. The target holder itself serves as a Faraday cup. A negative voltage of \unit[$-300$]{V} was applied to suppress secondary electrons in order to achieve a reliable charge collection. The areal particle density of Yb atoms of the different targets was measured to be between \unit[$0.76 \times 10^{18}$]{cm$^{-2}$} and  \unit[$1.49 \times 10^{18}$]{cm$^{-2}$} which translates to an areal density ranging from \unit[218]{$\mu$g/cm$^2$} to \unit[427]{$\mu$g/cm$^2$}. Taking into account the enrichment in $^{168}$Yb, this leads to an areal particle density of $^{168}$Yb target nuclei of \unit[$1.7 \times 10^{17}$]{cm$^{-2}$} to \unit[$5.2 \times 10^{17}$]{cm$^{-2}$}. The areal particle density was measured in steps of \unit[2]{mm} over the whole target to detect possible inhomogeneities, see Fig.~\ref{fig:rbs}. This example shows, that target inhomogeneities mount up to \unit[25]{\%} over the irradiated area. A similar pattern arises for the remaining targets. This is due to the target production process as all targets were produced in one step and thus did not have the same distance to the material that was evaporated. Hence, the target material was not evaporated homogeneously on the backing material and the maximum amount is not located at the center. Therefore, the mean thicknesses were calculated by the weighted average over the irradiated area and used for data analysis. The uncertainties in the number of target nuclei are between \unit[5]{\%} and \unit[7]{\%}. These were obtained by Gaussian error propagation using the systematic error of the RBS measurement and the statistical error of the calculation of the weighted average. The influence of the resulting inhomogeneous activity distribution on the detector efficiencies was investigated by means of a \textsc{Geant4} \cite{Geant4} simulation, see Sec.~\ref{sec:efficiency} for details. A second RBS measurement was performed after the $\gamma$-ray counting. This measurement confirmed, that no target material was lost during the experiment within the given uncertainties. 

\subsection{Experimental setup at PTB}
\begin{figure}[thb]
\centering
 \includegraphics[width=\textwidth]{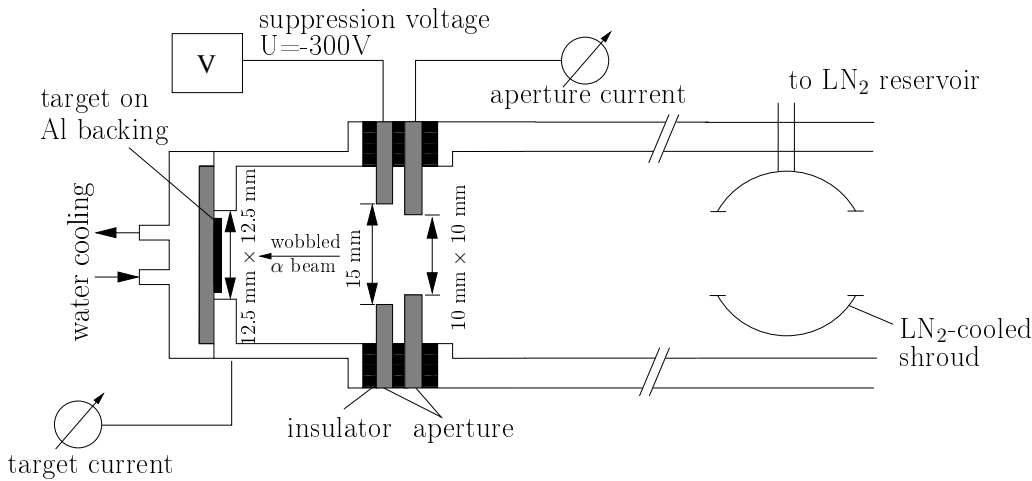}
\caption{Sketch of the activation chamber at PTB. The suppression voltage of $U=-$\unit[300]{V} is used to suppress secondary electrons in order to ensure a reliable charge collection. A LN$_2$-cooled shroud is used to reduce the buildup of carbon deposits on the target, scale changed for position of the shroud. Water cooling is applied to prevent an overheating of the target material.}
\label{fig:ptb}
\end{figure}

$^{4}$He$^{2+}$ ions were accelerated by the cyclotron and delivered to the targets which were placed inside a target chamber serving as a Faraday cup. Figure \ref{fig:ptb} shows a sketch of the experimental setup. The charge deposited on the target was measured by a current integrator. A scaler was used to record the beam current every \unit[60]{s}. This record was later used for a correction of current fluctuations and to determine the absolute number of $\alpha$ particles impinging on the target. The uncertainty in charge collection is \unit[1]{\%}. A negatively-charged diaphragm is utilized at the entrance of the activation chamber, where a voltage of $U=-$\unit[300]{V} is applied to suppress secondary electrons and thus to ensure a reliable charge collection. The $\alpha$ beam was wobbled over the target to guarantee a homogeneous illumination. Before each run, a quartz window was mounted at the target position to check the beam position and illumination. Water cooling was applied at the back of the target to prevent it from overheating. The beam spot had a rectangular shape with a size of about \unit[10]{mm}~$\times$~\unit[10]{mm}. Upstream the target chamber a LN$_2$-cooled trap was installed in order to reduce the buildup of carbon deposits on the target. The uncertainty given for the incident energy of the $\alpha$ particles $E_0$ is \unit[$\pm$25]{keV}. The incident energy of the $\alpha$ particles is determined by the field calibration of two analyzing magnets as well as by a time-of-flight measurement of the $\alpha$ particles \cite{Boettger}.

\subsection{$\gamma$-ray counting procedure}
\label{sec:gammacounting}
\begin{figure}[tb]
\centering
 \includegraphics[width=\textwidth]{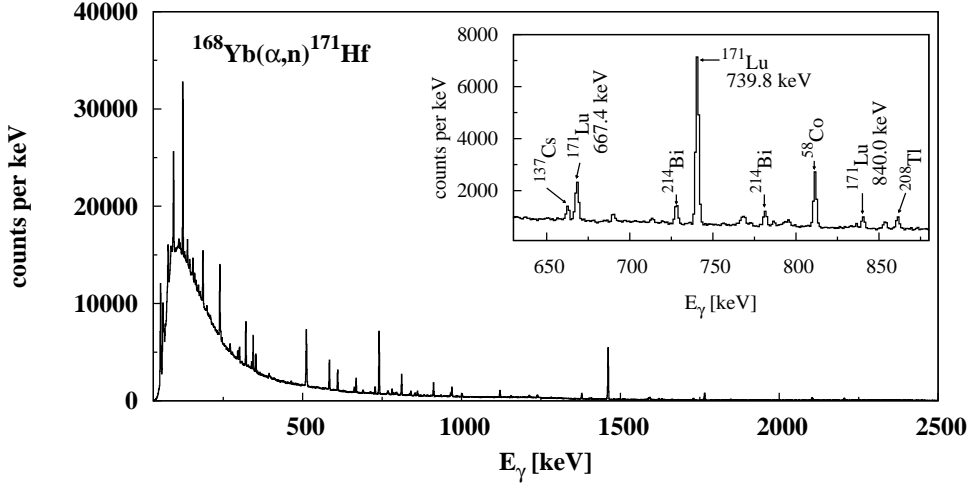}
\caption{Typical summed $\gamma$-ray spectrum taken with two HPGe clover detectors. The target was irradiated with \unit[15.1]{MeV} $\alpha$ particles. In the inset, the three transitions used for data analysis are highlighted. Additionally marked there are transitions stemming from natural background and reactions on target contaminants as the e.c. decay of $^{58}$Co. This spectrum was recorded for about 23 hours.}
\label{fig:spektrum_a_n}
\end{figure}

During the irradiation, both reactions took place simultaneously. The $\gamma$-ray counting was carried out with different types of HPGe detectors, due to the very different $\gamma$-ray energies of the decay of the produced unstable nuclei. The $^{168}$Yb($\alpha$,n) reaction produces the unstable reaction product $^{171}$Hf. The activities of the irradiated targets were measured at the Institute for Nuclear Physics at the University of Cologne using two clover-type HPGe detectors (referred to as clover setup), each with a relative efficiency of \unit[100]{\%} at \unit[$E_\gamma = 1.33$]{MeV} compared to a \unit[7.62]{cm}~$\times$~\unit[7.62]{cm} NaI detector. The detectors are shielded with a \unit[10]{cm} thick lead wall in order to suppress natural background. Furthermore, a copper sheet with a thickness of \unit[3]{mm} is used to shield X-rays stemming from the lead. In total, eight crystals are available, which in principle makes it possible to measure cross sections using effectively the $\gamma \gamma$ coincidence technique to suppress the background \cite{Sauerwein11}. However, this technique cannot be used in the present case, since the three strongest transitions are not emitted in a cascade. It was possible, however, to determine the cross sections using singles spectra.
As the absolute decay intensities of the electron capture of $^{171}$Hf are not known, the electron capture of $^{171}$Lu with a half-life of \unit[(8.24 $\pm$ 0.03)]{d} \cite{NNDC} was used to determine the cross sections. This decay leads to excited states in $^{171}$Yb which decay by emitting $\gamma$ rays.  The three strongest transitions with energies of \unit[$E_{\gamma,1} = 667.4$]{keV}, \unit[$E_{\gamma,2} = 739.8$]{keV}, and \unit[$E_{\gamma,3} = 840.0$]{keV} \cite{NNDC} were used to determine cross sections. Table~\ref{tab:decaydata} gives an overview of the relevant decay parameters. A typical $\gamma$-ray spectrum recorded by the clover setup from a target irradiated with $\alpha$ particles with an energy of \unit[$E_0 = 15.1$]{MeV} is shown in Fig.~\ref{fig:spektrum_a_n}. The inset in Fig.~\ref{fig:spektrum_a_n} shows a close-up view on the relevant energy region. The three transitions used for data analysis are marked by arrows. Additionally, significant peaks stemming from natural background and transitions stemming from reactions on target contaminants ($^{58}$Co) are marked. This spectrum was recorded over a period of 23 hours. The counting time varied for the other targets between 1 day for targets irradiated with higher beam energies and 4.5 days for targets irradiated with lower beam energies. 
In order to exclude systematic errors concerning the $\gamma$-ray counting procedure, the activity of three irradiated targets was additionally measured directly at PTB since this setup is well established. For this purpose a HPGe detector with a relative efficiency of \unit[70]{\%} was used (referred to as PTB detector). Within the given uncertainties, these independent measurements using the clover setup and the PTB detector gave consistent results.

\begin{figure}[thb]
\centering
 \includegraphics[width=\textwidth]{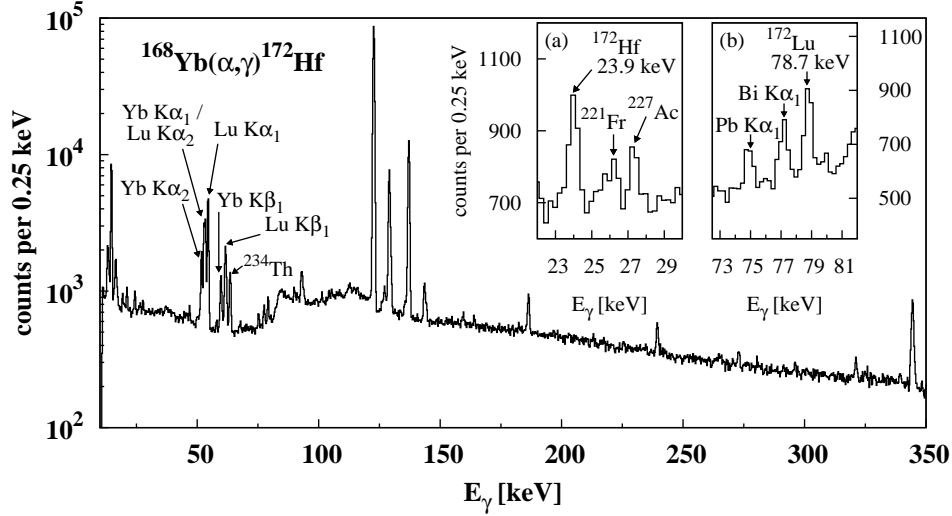}
\caption{X-ray and $\gamma$-ray spectrum taken with a LEPS detector for a target irradiated with \unit[$E_0~=~15.1$]{MeV}. In the total spectrum, the characteristic X-rays of Lu and Yb are marked with arrows, as well as the nearby $\gamma$-ray transition stemming from natural occurring $^{234}$Th. The insets (a) and (b) show close-up views around the energy regions of interest. The relevant $\gamma$-ray transitions with \unit[$E_\gamma = 23.9$]{keV} and \unit[$E_\gamma = 78.7$]{keV} are marked with arrows. Additionally, $\gamma$-ray transitions from natural occurring background as well as the emitted X-rays of Bi and Pb are indicated. This spectrum was recorded for about 14 days.}
\label{fig:spektrum_a_g}
\end{figure}

By the $^{168}$Yb($\alpha$,$\gamma$) reaction the unstable reaction product $^{172}$Hf was produced which decays with a half-life of \unit[(1.87 $\pm$ 0.03)]{a} \cite{NNDC} via electron capture. The counting of the low-energy $\gamma$ rays was carried out using a Canberra type GL2015R LEPS (Low Energy Photon Spectrometer) at the Institute for Nuclear Research (ATOMKI) in Debrecen, Hungary \cite{Kiss11_1}. The measurement was carried out at a distance of \unit[1]{cm} between the target and detector end cap.
This detector is equipped with an almost 4$\pi$ shielding consisting of inner layers of copper and cadmium as well as an \unit[8]{cm} thick outer lead layer. With this detector it was possible to observe the low-energy $\gamma$-ray transition with an energy of \unit[$E_\gamma = 23.9$]{keV} \cite{NNDC} of the $\gamma$ decay of $^{172}$Hf which could be used to determine the cross sections of the ($\alpha$,$\gamma$) reaction.
The decay of $^{172}$Lu with a half-life of \unit[(6.70~$\pm$~0.03)]{d} leads to $^{172}$Yb, where a photon with an energy of \unit[$E_\gamma = 78.7$]{keV} is emitted \cite{NNDC}. This decay was additionally used to determine the ($\alpha$,$\gamma$) cross sections. Figure~\ref{fig:spektrum_a_g} shows a $\gamma$-ray spectrum recorded for about 14 days with the LEPS detector after irradiation with $\alpha$ particles with an incident energy of \unit[15.1]{MeV}. The insets (a) and (b) show the spectrum focused on the energy regions of interest, \textit{i.e.} around \unit[$E_\gamma = 23.9$]{keV} and \unit[$E_\gamma = 78.7$]{keV}, respectively. The transitions that are used for cross-section determination are marked by arrows, as well as the emitted X-rays of Lu and Yb. Moreover, $\gamma$-ray transitions stemming from natural background are indicated by arrows as well as the K$\alpha_1$-transitions of Bi and Pb in Fig.~\ref{fig:spektrum_a_g}~(b). The counting periods varied between 13 days and 19 days.

\begin{table}[tb]
\centering
\caption{Listed are the decay parameters of the reaction products and their decay products that were used for data analysis. Data taken from \cite{NNDC}.}
\begin{tabular}{c c c c c}

Reaction 			& Isotope 	& Half-life 	      & $E_\gamma$ [keV]     & $I_\gamma$          \\ \hline 
\vspace{-0.3cm}
& & & & \\
$^{168}$Yb($\alpha$,$\gamma$)	& $^{172}$Hf	& \unit[(1.87 $\pm$ 0.03)]{a} & 23.9  & 0.203 $\pm$ 0.017 \\
$^{168}$Yb($\alpha$,$\gamma$)	&$^{172}$Lu 	& \unit[(6.70 $\pm$ 0.03)]{d} & 78.7  & 0.106 $\pm$ 0.005 \\
$^{168}$Yb($\alpha$,n)		& $^{171}$Lu	& \unit[(8.24 $\pm$ 0.03)]{d} & 667.4 & 0.111 $\pm$ 0.003 \\
$^{168}$Yb($\alpha$,n)		& $^{171}$Lu	& \unit[(8.24 $\pm$ 0.03)]{d} & 739.8 & 0.479 $\pm$ 0.011   \\
$^{168}$Yb($\alpha$,n)		& $^{171}$Lu	& \unit[(8.24 $\pm$ 0.03)]{d} & 840.0 & 0.030 $\pm$ 0.001 \\ \hline
\end{tabular}
\label{tab:decaydata}
\end{table}

\section{Data analysis}
\label{sec:dataanalysis}
In the case of the ($\alpha$,n) reaction, the product $^{171}$Hf decays to $^{171}$Lu, which is also unstable and decays further to the stable isotope $^{171}$Yb. As the latter decay was used for data analysis, the absolute number of decays of $^{171}$Lu must be used. The absolute number of $^{171}$Lu nuclei $\Delta N$ decaying during the counting period is related to the counts measured in the full-energy peak $Y(E_\gamma)$ at an energy $E_\gamma$ as follows:
\begin{equation}
\label{eq:yield}
\Delta N = \frac{Y(E_\gamma)}{I_\gamma(E_\gamma) \, \varepsilon(E_\gamma) \, \tau} \, \mathrm{,}  
\end{equation}
where $I_\gamma(E_\gamma)$ denotes the absolute $\gamma$-ray intensity of a given transition with energy $E_\gamma$ and $\varepsilon(E_\gamma)$ the absolute full-energy peak efficiency at this energy. The parameter $\tau = t_\mathrm{{Live}}/t_\mathrm{{Real}}$ accounts for the dead time correction of the data acquisition system which was found to be of the order of \unit[1]{\%} or less. From the radioactive decay law, $\Delta N$ can also be derived using the amount of $^{171}$Hf  $N_{\mathrm{act}}^{\mathrm{Hf}}$ and $^{171}$Lu $N_{\mathrm{act}}^{\mathrm{Lu}}$ nuclei at the end of the activation period. The obtained equations are solved for small intervals of \unit[$\Delta t_i \approx 60$]{s}. Within these intervals, the production rate $P_i$ is assumed to be constant. This production rate is proportional to the number of impinging projectiles $N_\alpha$. During the activation period, given by $M \, \Delta t_i$, an amount of $N_{\mathrm{Prod}}$ $^{171}$Hf nuclei are produced. A part of the $^{171}$Hf nuclei decay during the irradiation. The absolute number $N_{\mathrm{act}}^{\mathrm{Hf}}$ at the end of the activation period can be derived from the experimentally known $\Delta N$, which is given by

\begin{equation}
 \begin{split}
  N_{\mathrm{act}}^{\mathrm{Hf}} = & f_\mathrm{act} N_\mathrm{prod} \, ,
 \end{split}
\end{equation}
where $f_\mathrm{act}$ is given by

\begin{equation}
 \begin{split}
  f_\mathrm{act} =& \dfrac{\left(1 - e^{-\lambda_\mathrm{Hf} \, \Delta t_i} \right)}{\lambda_\mathrm{Hf}} \dfrac{\sum_{i=1}^M P_i \, e^{-\lambda_\mathrm{Hf} \left(M-i\right)\Delta t_i }}{\sum_{i=1}^M P_i \Delta t_i} 
 \end{split}
\end{equation}

The quantity $\lambda_{\mathrm{Hf}}$ denotes the decay constant of $^{171}$Hf. From this amount of $^{171}$Hf nuclei produced, the cross section can finally be determined by

\begin{equation}
\label{eq:crosssection}
\sigma\left(E_0\right) = \frac{N_{\mathrm{prod}}}{N_\mathrm{target}\, N_\alpha}\, \mathrm{,}
\end{equation}

where $\sigma\left(E_0 \right)$ denotes the reaction cross section at an $\alpha$-particle energy $E_0$ and $N_\mathrm{target}$ stands for the areal particle density of target nuclei. The same procedure to determine the number of reaction products also holds for the ($\alpha$,$\gamma$) reaction, where the decay of $^{172}$Lu was used to determine the total cross section. 
For the case, that only one decay is taken into account, such as for the e.c. decay of $^{172}$Hf, the standard case of an activation experiment is present. See, $e. \, g.$, Ref.~\cite{Sauerwein11} for a derivation of the factor $f_{\mathrm{act}}$.

\subsection{Detector efficiencies}
\label{sec:efficiency}
\begin{figure}[tb]
\centering
 \includegraphics[width=\textwidth]{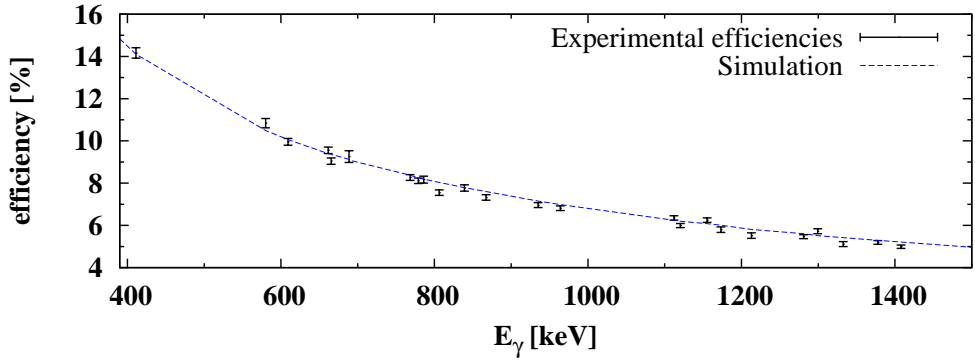}
\caption{(Color online) The experimental efficiencies without summing effects are compared to the \textsc{Geant4} simulation for the clover setup in close geometry. Depicted is the summed efficiency of both clover detectors. The experimental efficiencies are well described by the simulation. The experimental data was obtained by scaling the efficiencies measured at the far geometry using a conversion factor, see text. Only the energy region relevant for the experiment is shown here.}
\label{fig:efficiency_clover}
\end{figure}

In order to determine reaction cross sections the absolute full-energy peak efficiencies of all detectors have to be known. The general procedure to determine the absolute full-energy peak efficiencies was the same for every detector. To account for summing effects of the calibration sources, the detector efficiencies were measured in a far geometry in a first step. In a second step, the efficiencies were measured in a close geometry using a calibration source, where no $\gamma$ rays are emitted in a cascade and a conversion factor was determined to subsequently scale the measured efficiencies at the far geometry. For the Cologne clover setup and the PTB detector this was accomplished using a $^{137}$Cs source. Monte Carlo simulations with \textsc{Geant4} were performed and showed a very good reproduction of the experimental efficiencies without summing effects, see Fig. \ref{fig:efficiency_clover} for an example. As stated in Sec. \ref{subsec:target}, the target material was not distributed homogeneously. As the efficiency calibration was performed using point-like sources, the extended geometry of the target and the observed inhomogeneities must be taken into account. This was accomplished by a further \textsc{Geant4} simulation. The simulated efficiencies using an extended inhomogeneous target geometry agree within less than 1 \% with the simulated efficiencies using a point-like geometry.

For the LEPS detector a self-produced $^{131}$Cs source, that was produced with the $^{127}$I($\alpha$,$\gamma$) reaction \cite{Kiss12}, was used for this purpose, since this nucleus decays only via X-ray emission with an energy of  \unit[$E_{K \alpha 1 / 2} \approx 29.6$]{keV}. In addition, for the efficiency at \unit[$E_\gamma~=~$78.7]{keV} a $^{172}$Lu source was produced via the $^{169}$Tm($\alpha$,n) reaction. As the reaction product $^{172}$Hf decays via $\gamma$-ray cascades, the obtained efficiencies were corrected for the coincidence-summing effect as given in Ref.~\cite{Semkov90}. The efficiency uncertainty is higher than usual for this setup, see $e.g.$ \cite{Kiss11_1}, because the irradiated area is larger than the ones usually used for this setup. However, Monte Carlo simulations showed, that efficiency curves at positions varied by up to \unit[5]{mm} in each direction agreed within less than \unit[5]{\%}. This value enters the efficiency uncertainty according to Gaussian error propagation and embodies the major source of uncertainty for the efficiency determination. The $\gamma$-attenuation factor was found to be less than \unit[1]{\%} for \unit[23.8]{keV} and less than \unit[0.5]{\%} for \unit[78.7]{keV} using the LISE code \cite{Lise} and therefore it was neglected.

An overview about the used calibration sources and distances concerning the efficiency calibration can be found in Table~\ref{tab:efficiency}.

\begin{table}[tb]
\centering
\caption{Distances and used calibration sources to determine the absolute full-energy peak efficiencies. A scaling factor was determined between the close counting distance and the far geometry to account for summing effects. The absolute efficiency was determined using calibrated sources, which are listed in the rightmost column, except for the self-produced $^{131}$Cs and $^{172}$Lu source, respectively. Details are given in the text. }
\begin{tabular}{c c c c}
Setup & Counting distance [mm] & Far geometry [mm] &  Used calibration sources\\ \hline
\vspace{-0.2cm}
& & &\\

Clover & 13 & 100 & $^{57}$Co, $^{60}$Co, $^{133}$Ba,  \\
& & & $^{137}$Cs, $^{152}$Eu, $^{226}$Ra \\
PTB & 15 & 70 & $^{60}$Co, $^{137}$Cs, $^{152}$Eu \\
LEPS & 10 & 100 & $^{57}$Co, $^{133}$Ba, $^{152}$Eu, $^{241}$Am,  \\
& & & $^{131}$Cs (rel.), $^{172}$Lu (rel.) \\
 \hline

\end{tabular}
\label{tab:efficiency}
\end{table}

\section{Results and discussion}
\label{sec:results}
\subsection{Experimental cross sections of the reactions $^{168}$Yb($\alpha$,n)$^{171}$Hf and $^{168}$Yb($\alpha$,$\gamma$)$^{172}$Hf}
\label{sec:results_an}

Tables~\ref{tab:results_an} and \ref{tab:results_ag} show the experimentally determined cross sections for the ($\alpha$,n) and ($\alpha$,$\gamma$) reaction, respectively. The results shown here were obtained by calculating the weighted average of the cross-section results for each $\gamma$-ray transition mentioned in Sec.~\ref{sec:dataanalysis} for the respective nucleus. As stated in Sec.~\ref{sec:gammacounting}, for the ($\alpha$,n) reaction the activity of three targets irradiated with incident $\alpha$-particle energies of \unit[15.10]{MeV}, \unit[14.55]{MeV}, and \unit[13.76]{MeV} was additionally measured at PTB. Both measurements yielded consistent results within the given uncertainties. 

In Table \ref{tab:results_ag} the experimental results for the reaction $^{168}$Yb($\alpha$,$\gamma$)$^{172}$Hf are presented. For this reaction, cross sections for five incident $\alpha$-particle energies between \unit[13.5]{MeV} and \unit[15.10]{MeV} could be determined using two different electron capture decays, which yielded consistent results. For comparison with the theoretical calculations, the weighted average of these results were used. For the target irradiated with the lowest incident energy of \unit[$E_0~=~12.9$]{MeV}, the induced activity was too low to determine the cross section using the LEPS detector. Nevertheless, it was possible to determine an upper limit of \unit[0.011]{mb}.

The procedure to correct the incident $\alpha$-particle energy with respect to energy loss and straggling inside the target material follows the one of Ref.~\cite{Sauerwein11}. In the present case, the energy loss of the $\alpha$ particles ranged from \unit[34]{keV} to \unit[75]{keV}, whereas the straggling inside the target material ranged from \unit[9]{keV} to \unit[15]{keV}.

For the energy of \unit[$E_{\mathrm{c.m.}}=14.2$]{MeV}, the cross section was also obtained using the X-ray counting method as presented in Ref.~\cite{Kiss11}. The disadvantage of this method is that it is impossible to distinguish between X-rays emitted by different isotopes. Thus, the half-lives of the isotopes must be vastly different. The reaction products of $\alpha$-induced reactions on isotopes besides $^{168}$Yb within the target are either stable or have half-lives which are short compared to the waiting period between irradiation and counting. The only exception is $^{175}$Hf, the product of the $^{172}$Yb($\alpha$,n) or $^{171}$Yb($\alpha$,$\gamma$) reaction. This isotope has a half-life of \unit[(70~$\pm$~2)]{d} \cite{NNDC}. In order to disentangle the amount of emitted X-rays stemming from the electron capture decay of $^{175}$Hf, the $\gamma$-ray transition with \unit[$E_\gamma=343.4$]{keV} originating from this decay was used. It was possible to determine the cross section for this energy using the K$\alpha_1$ transition following the e.c. decay of $^{172}$Hf. This X-ray has an energy of \unit[54.07]{keV} and an absolute intensity of \unit[(63~$\pm$~6)]{\%} \cite{NNDC}. The remaining X-rays could not be used, since they cannot be separated in the spectrum, see Fig.~\ref{fig:spektrum_a_g}. With this procedure, a cross section of \unit[(0.16~$\pm$~0.05)]{mb} was determined, which is in excellent agreement with the cross sections obtained from $\gamma$ counting, see Table~\ref{tab:results_ag}.

One has to note here, that the presented cross sections are not corrected for electron screening \cite{Rolfs88,Rolfs06}. The screening potential for the present case is \unit[$U_e=20.6$]{keV}, when the appropriate charge scaling of Ref.~\cite{Kettner06} is applied. This would lead to a decrease of the measured cross section of \unit[5]{\%} to \unit[6]{\%}, depending on the energy.

\begin{table}[tb]
\centering
\caption{Summary of experimental cross sections for each center-of-mass energy $E_{\mathrm{c.m.}}$ for the $^{168}$Yb($\alpha$,n)$^{171}$Hf reaction. The weighted average, if two or more results were available, is shown in the rightmost column. The uncertainty is given by the variance of the weighted mean. The detectors used are also indicated, as well as the areal particle density $m$ of $^{168}$Yb target nuclei.}
\begin{tabular}{c c c c c}
$E_{\mathrm{c.m.}}$ [keV]	& $m$ [$\mathrm{cm}^{-2}$]		&	Detector	&  $\sigma$ [mb] &	$\bar{\sigma}$ [mb]	\\ \hline
\vspace{-0.2cm}
& & &\\
12534 $\pm$ 28		& (4.93 $\pm$ 0.28) $\times \, 10^{17}$	&	clover setup	& 0.012 $\pm$ 0.001	&	$-$\\
13155 $\pm$ 28		& (1.70 $\pm$ 0.09) $\times \, 10^{17}$	&	clover setup	& 0.045 $\pm$ 0.005	&	$-$\\
13406 $\pm$ 28		& (4.99 $\pm$ 0.28) $\times \, 10^{17}$	&	clover setup	& 0.066 $\pm$ 0.007	&	0.068 $\pm$ 0.007\\
			& 					&	PTB		& 0.070 $\pm$ 0.007	&	\\
13835 $\pm$ 28		& (5.23 $\pm$ 0.27) $\times \, 10^{17}$	&	clover setup	& 0.160 $\pm$ 0.015	&	$-$\\
14178 $\pm$ 28		& (4.86 $\pm$ 0.37) $\times \, 10^{17}$	&	clover setup	& 0.35 $\pm$ 0.04	&	0.37 $\pm$ 0.04\\
			&					&	PTB		& 0.38 $\pm$ 0.04	&	\\
14732 $\pm$ 26		& (2.67 $\pm$ 0.12) $\times \, 10^{17}$	&	clover setup	& 1.28 $\pm$ 0.12	&	1.29 $\pm$ 0.12\\
			&					&	PTB		& 1.30 $\pm$ 0.12	&	\\ \hline
\end{tabular}
\label{tab:results_an}
\end{table}

\begin{table}[tb]
\centering
\caption{Summary of experimental cross sections for each center-of-mass energy $E_{\mathrm{c.m.}}$ for the $^{168}$Yb($\alpha$,$\gamma$)$^{172}$Hf reaction. The weighted average, if two or more results were available, is shown in the rightmost column. The uncertainty is given by the variance of the weighted mean. The induced activity was measured using a LEPS detector. Also noted is the corresponding nucleus whose decay is considered and the areal particle density of $^{168}$Yb nuclei.}
\begin{tabular}{c c c c c c}
$E_{\mathrm{c.m.}}$ [keV] & $m$ [$\mathrm{cm}^{-2}$] &	Nucleus 		&  $\sigma$ [mb]  &	$\bar{\sigma}$ [mb]\\  \hline
\vspace{-0.2cm}
& & & \\
12534 $\pm$ 28	& (4.93 $\pm$ 0.28) $\times \, 10^{17}$	&				& < 0.011 $\pm$ 0.003 	&	$-$	\\
13155 $\pm$ 28	& (1.70 $\pm$ 0.09) $\times \, 10^{17}$&	$^{172}$Hf		& 0.028 $\pm$ 0.005&	0.028 $\pm$ 0.008	\\
		& 					&	$^{172}$Lu		& 0.029 $\pm$ 0.010&		\\
13406 $\pm$ 28	& (4.99 $\pm$ 0.28) $\times \, 10^{17}$&	$^{172}$Hf		& 0.059 $\pm$ 0.010&	0.059 $\pm$ 0.011	\\
		& 					&	$^{172}$Lu		& 0.058 $\pm$ 0.014&		\\
13835 $\pm$ 28	& (5.23 $\pm$ 0.27) $\times \, 10^{17}$&	$^{172}$Hf		& 0.095 $\pm$ 0.017&	0.098 $\pm$ 0.019 	\\
		& 					&	$^{172}$Lu		& 0.11 $\pm$ 0.03& 		\\
14178 $\pm$ 28	&(4.86 $\pm$ 0.37) $\times \, 10^{17}$	 &	$^{172}$Hf		& 0.15 $\pm$ 0.02&	0.15 $\pm$ 0.02	\\
		& 					&	$^{172}$Lu		& 0.16 $\pm$ 0.03&		\\
		&					&	$^{172}$Lu K$\alpha_1$	& 0.16 $\pm$ 0.05&		\\
14732 $\pm$ 26	& (2.67 $\pm$ 0.12) $\times \, 10^{17}$&	$^{172}$Hf		& 0.29 $\pm$ 0.04&	0.29 $\pm$ 0.04	\\
		&					 &	$^{172}$Lu		& 0.27 $\pm$ 0.05&		\\ \hline
\end{tabular}
\label{tab:results_ag}
\end{table}

\subsection{Comparison with statistical model calculations}
\label{sec:theory}
In order to enable a comparison with theoretical calculations, the astrophysical S factors were calculated using the weighted averages of the cross sections. The results of the S factors are given in Table~\ref{tab:sfactor} for both reactions.

\begin{table}[tb]
\centering
\caption{Astrophysical S factors as a function of center-of-mass energies for the $^{168}$Yb($\alpha$,n) reaction (middle column) and $^{168}$Yb($\alpha$,$\gamma$) reaction (right column). These were derived from the weighted average of the cross sections.}
\begin{tabular}{c c c}
& $^{168}$Yb($\alpha$,n) & $^{168}$Yb($\alpha$,$\gamma$) \vspace{2pt} \\
$E_{\mathrm{c.m.}}$ [keV] & S factor [$10^{29}$ MeVb]  	& S factor [$10^{29}$ MeVb]  	\\  \hline
\vspace{-0.2cm}             
& &  \\
12534 $\pm$ 28	& 5.94 $\pm$ 0.62 	&	< 5.59 $\pm$ 1.46	\\
13155 $\pm$ 28	& 3.65 $\pm$ 0.37	&	2.25 $\pm$ 0.55	\\
13406 $\pm$ 28	& 2.74 $\pm$ 0.27	&	2.37 $\pm$ 0.45	\\
13835 $\pm$ 28	& 2.07 $\pm$ 0.18	&	1.27 $\pm$ 0.25	\\
14178 $\pm$ 28	& 1.99 $\pm$ 0.22	&	0.83 $\pm$ 0.13	\\
14732 $\pm$ 26	& 1.83 $\pm$ 0.17	&	0.40 $\pm$ 0.06	\\ \hline

\end{tabular}
\label{tab:sfactor}
\end{table}

As stated in Sec.~\ref{sec:motivation}, the cross section of the ($\alpha$,n) reaction is also strongly dependent on the other nuclear ingredients besides the $\alpha$ width. This is even more evident in the case of the ($\alpha$,$\gamma$) reaction at energies above the neutron emission threshold at \unit[$\approx 12$]{MeV}, \textit{i.e.} inside the measured energy range. Nevertheless, the aim of the following procedure was to find the best theoretical description to account for both reactions simultaneously.

\begin{figure}[tb]
\centering
 \includegraphics[width=\textwidth]{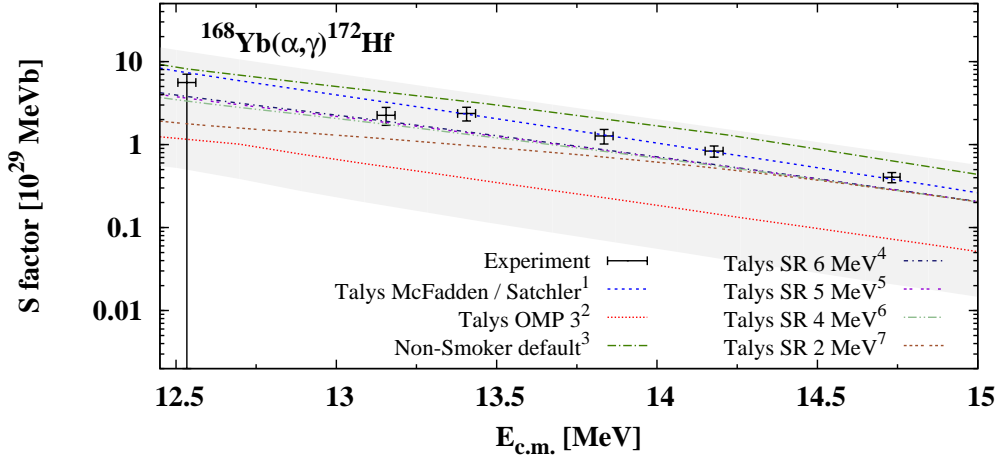}
\caption{(Color online) Astrophysical S factors of the $^{168}$Yb($\alpha$,$\gamma$)$^{172}$Hf reaction as a function of center-of-mass energies. These are compared to theoretical predictions from the \textsc{talys} \cite{Talys} and \textsc{non-smoker} \cite{Nonsmoker} code. The gray shaded area was obtained by varying the different input parameter combinations of the \textsc{talys} code only. In order to achieve a good description of the experimental data, the photon strength functions, nuclear level densities, and neutron-OMPs were varied as well. The $\alpha$-OMPs as labeled in the figure were combined with the following input parameters: (1) photon strength function of Ref.~\cite{Brink57} (Brink/Axel), nuclear level density of Ref.~\cite{Talys} (generalized superfluid), and neutron-OMP of Ref.~\cite{Koning03} (Koning/Delaroche). (2) photon strength function of Ref.~\cite{Goriely04} (microscopic), nuclear level density of Ref.~\cite{Goriely08} (microscopic), and neutron-OMP of Ref.~\cite{Bauge01} (semi-microscopic). (3) $\alpha$-OMP of Ref.~\cite{McFadden66} (McFadden/Satchler), photon strength function of Ref.~\cite{Nonsmoker} (standard Lorentzian with modified low-energy tail), nuclear level density of Ref.~\cite{Rauscher97} (based on shifted Fermi gas), and neutron-OMP of Ref.~\cite{Jeukenne77} (based on microscopic nuclear matter calculations). (4)-(7) same as (1), but with a modified $\alpha$-OMP of Ref.~\cite{Sauerwein11} with \unit[$a_E=6$]{MeV}, \unit[5]{MeV}, \unit[4]{MeV}, and \unit[2]{MeV}.}

\label{fig:talys_ag}
\end{figure}

\begin{figure}[tb]
\centering
 \includegraphics[width=\textwidth]{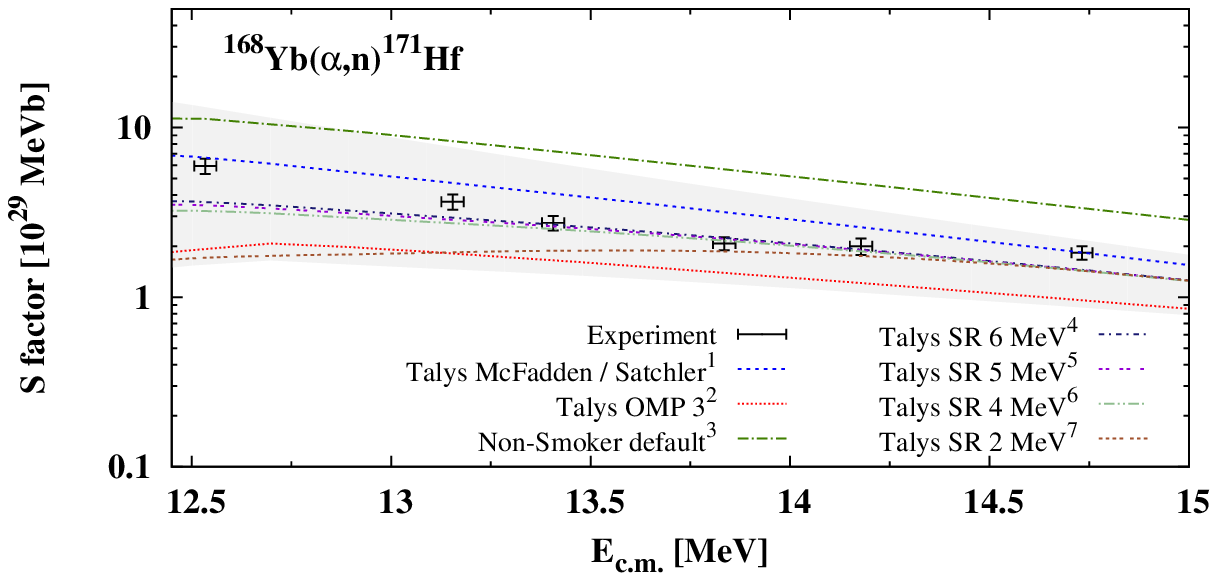}
\caption{(Color online) Same as Fig.~\ref{fig:talys_ag}, but for the $^{168}$Yb($\alpha$,n)$^{171}$Hf reaction.} 
\label{fig:talys_an}
\end{figure}

The \textsc{talys} 1.4 code \cite{Talys}, which was used for theoretical calculations, provides a variety of input parameters for HF calculations. These include different phenomenological and microscopic descriptions of $\alpha$-OMPs \cite{Watanabe58,McFadden66,Demetriou02}, photon strength functions \cite{Brink57,Kopecky90,Goriely02_1,Goriely04}, nuclear level densities \cite{Talys,Demetriou01,Goriely08,Koning08}, and neutron-OMPs \cite{Koning03,Bauge01,Goriely07}. All of these models can be found the Reference Input Parameter Library (RIPL3) \cite{Capote09}. Due to the quite pronounced sensitivities of both cross sections to the $\alpha$ width, $\gamma$ width, and neutron width, it is important to estimate, which influences the different combinations of nuclear physics input parameters have on the cross sections. This was done by investigating, which range of cross sections can be reached using the different combinations of ingredients available in \textsc{talys}. For this purpose, two restrictions were made on the choice of the input parameters. Firstly, mixed microscopic and phenomenological combinations of nuclear level densities and photon strength functions were excluded. Secondly, only those microscopic combinations of nuclear level densities and photon strength functions that are calculated within the same theoretical framework were used. This is the case for the nuclear level density of Ref.~\cite{Demetriou01} combined with a photon strength function of Ref.~\cite{Goriely02_1} which are calculated within the framework of the Hartree-Fock-BCS model. Similarly, the combination of the combinatorial nuclear level density of Ref.~\cite{Goriely08} and the photon strength function of Ref.~\cite{Goriely04} are both based on the Hartree-Fock-Bogolyubov model. The gray shaded area in Figs.~\ref{fig:talys_ag} and \ref{fig:talys_an} depicts the range spanned by the S factors obtained with the aforementioned combinations. This area points out, that the calculated S factor is to a large extent sensitive to combinations of different input parameters and that it is not sufficient to vary only the $\alpha$-OMP. Note, that the calculation using the \textsc{non-smoker} code \cite{Nonsmoker} in its default settings was not involved in the determination of this area. 

In a next step, from all these combinations out of this region the best description of the experimental data had to be found simultaneously for both reactions. First, the average deviation between the experimental and theoretical data was calculated for each energy. Additionally, this deviation was checked for constancy to account for a correct energy dependence of the model prediction. Secondly, the difference between the experimental values and the average deviation from the first step was investigated to assess the statistical scattering of the experimental values around the calculated ones. One has to note here, that there is no combination, which is the best one for all criteria described above for both reactions. Therefore, the best possible compromise was chosen in order to describe both reactions with one set of input parameters satisfactorily. In order to judge the goodness of the theoretical description, a $\chi^2$-value is also given for each input-parameter combination.

It was found, that the best description is given by the $\alpha$-OMP of Ref.~\cite{McFadden66} (McFadden/Satchler) combined with a phenomenological description of the photon strength function of Ref.~\cite{Brink57} (Brink/Axel), nuclear level density of Ref.~\cite{Talys} (Generalized Superfluid), and a phenomenological spherical neutron-OMP of Ref.~\cite{Koning03} (Koning/Delaroche). Figures \ref{fig:talys_ag} and \ref{fig:talys_an} show a comparison of the experimental S factor for the ($\alpha$,$\gamma$) and ($\alpha$,n) reaction and theoretical calculations. The reproduction of the experimental data of the ($\alpha$,$\gamma$) reaction is excellent using these parameters ($\chi^2=0.31$). Concerning the ($\alpha$,n) reaction, the experimental values are systematically overestimated by these parameters except for the data point at the highest energy ($\chi^2=1.26$). The McFadden/Satchler $\alpha$-OMP was obtained by fitting a Woods-Saxon potential to experimental elastic $\alpha$-scattering data at an energy of \unit[24.7]{MeV}. Scattering and reaction data at higher energies are successfully reproduced, but it often fails to describe data at lower energies, see, $e.g.$, \cite{Sauerwein11}. The present measurement was performed at energies considerably higher than the astrophysically relevant energy range for the ($\alpha$,$\gamma$) reaction. Therefore, the extrapolation down to lower energies, $i.e.$ inside the Gamow window, might not be reliable, although the experimental data at higher energies is reproduced well. 

The Brink/Axel parameterization of the photon strength function of Ref.~\cite{Brink57} as implemented in \textsc{talys} is obtained from systematics of the giant dipole resonance (GDR) and not adjusted for energies below $\approx$~\unit[8]{MeV}. Additionally, it has been shown, that this photon strength function is only applicable for $\gamma$-ray energies well above \unit[1]{MeV} to \unit[2]{MeV}, see \cite{Capote09} and references therein. However, $\gamma$ rays emitted with an energy of about \unit[3]{MeV} to \unit[4]{MeV} have the largest impact on the calculated cross section and the influence of low-energy $\gamma$ rays is negligible. Furthermore, the GDR in $^{172}$Hf is located at $E_{\mathrm{GDR}}=$\unit[14.3]{MeV} with a width of $\Gamma_{\mathrm{GDR}}=$\unit[4.2]{MeV}. Thus the range of excitation energies in this experiment significantly overlaps with the range of the GDR in $^{172}$Hf. Hence, in the present case, the standard Lorentzian parameterization is suitable to describe the photon strength function.

Ultimately, one is interested in a global description of the $\alpha$-OMP which is independent from locally adjusted parameters. Therefore, the experimentally obtained S factors are additionally compared to a calculation including a global semi-microscopic $\alpha$-OMP (OMP 3) of Ref.~\cite{Demetriou02} combined with a microscopic description of the other input parameters. For the photon strength function the microscopic calculation of Ref.~\cite{Goriely04} was used, where the $E1$ component was calculated within the scope of the Hartree-Fock-Bogolyubov (HFB) model. Furthermore the nuclear level density of Ref.~\cite{Goriely08} as well as a spherical semi-microscopic description of the neutron-OMP of Ref.~\cite{Bauge01} was used. This combination of nuclear models was chosen, because the photon strength function and nuclear level density are both calculated consistently within the same theoretical framework. It can be seen from Fig.~\ref{fig:talys_ag}, that the experimental data are underestimated by a factor of five ($\chi^2=24.85$). However, this input parameter combination yields a satisfactory reproduction of the overall energy dependence. The overall energy dependence of the ($\alpha$,n) reaction is described well, but the absolute values are underestimated significantly also in this case, see Fig.~\ref{fig:talys_an}. The real part of the used $\alpha$-OMP is obtained using a double-folding approach of Ref.~\cite{Kobos84}. The imaginary part is composed of a damped surface potential and a volume part whose depth and geometry parameters are fitted to the bulk of existing experimental data at low energies, $i.e.$, close at or below the Coulomb barrier. OMP 3, in addition, is a dispersive optical model potential. Furthermore, the surface absorption decreases rapidly with higher energies which is accounted for by an exponential damping in the surface potential. Hence, the imaginary part of OMP 3 is composed of a damped surface potential and a volume part, for details see Ref.~\cite{Demetriou02}.

A comparison with calculations of the widely used \textsc{non-smoker} code in its default settings (Non-Smoker default) \cite{Nonsmoker} is also given. This calculation uses the phenomenological $\alpha$-OMP of Ref.~\cite{McFadden66}. The comparison with the ($\alpha$,$\gamma$) S factors in Fig.~\ref{fig:talys_ag} shows, that the experimental data is slightly overpredicted, but the energy dependence is very well reproduced ($\chi^2=3.06$). Concerning the ($\alpha$,n) reaction, the energy dependence is also well described, but the absolute value is significantly overpredicted ($\chi^2=12.48$). 

It was shown recently in Ref.~\cite{Rauscher12_1}, that a modification of the Sauerwein/Rauscher $\alpha$-OMP from Ref.~\cite{Sauerwein11} is able to reproduce experimental data of the $^{169}$Tm($\alpha$,$\gamma$)$^{173}$Lu and $^{169}$Tm($\alpha$,n)$^{172}$Lu reactions. Motivated by this success, the present data are compared to the Sauerwein/Rauscher potential using the same modifications. The phenomenological $\alpha$-OMP from Ref.~\cite{Sauerwein11} is a modification of the McFadden/Satchler $\alpha$-OMP. The depth of the real part and the geometry of the real and imaginary part was retained, but the strength of the volume imaginary part $W$ was made energy-dependent:

\begin{equation}
 \unit[W = \frac{25}{1+e^{\left(0.9 E_C - E_{\mathrm{c.m.}}\right)/a_E}}]{MeV} \, ,
\end{equation}
  
where $E_C$ denotes the energy of the Coulomb barrier and $a_E$ is the diffuseness of the Fermi-type function. This $\alpha$-OMP transforms into the McFadden/Satchler for higher energies, where experimental data are often successfully described. For the case of the $^{141}$Pr($\alpha$,n)$^{144}$Pm reaction, no modification of the parameter $a_E$ was necessary \cite{Sauerwein11}. However, for the present investigated $\alpha$-induced reactions on $^{168}$Yb, the energy dependence is too steep when going to low energies, yielding $\chi^2$-values of $\chi^2=12.95$ for the ($\alpha$,n) and $\chi^2=3.78$ for the ($\alpha$,$\gamma$) reaction, respectively. Thus, varied values of $a_E = $ \unit[4, 5, and 6]{MeV} were used. Figures~\ref{fig:talys_ag} and \ref{fig:talys_an} show, that calculations using the Sauerwein / Rauscher $\alpha$-OMP with varied values of $a_E = $ \unit[4, 5, and 6]{MeV} yield an excellent reproduction of the energy dependence over the whole energy region simultaneously for both reactions. The absolute experimental values are slightly underpredicted, but still the difference is less than a factor of 2. The best model prediction with this input parameters is given by the one using $a_E = $ \unit[6]{MeV}, which gives $\chi^2$-values of $\chi^2=1.77$ for the ($\alpha$,n) and $\chi^2=1.21$ for the ($\alpha$,$\gamma$) reaction, respectively.

Although the sensitivity of the cross sections of both reactions to input parameters besides the $\alpha$-OMP is rather high, one can draw conclusions on the $\alpha$-OMP. The global semi-microscopic $\alpha$-OMP of Ref.~\cite{Demetriou02} (OMP 3) is not able to reproduce the experimental data in the present case, if combined with any other description of photon strength function, nuclear level density, and neutron-OMP. On the other hand, adequate descriptions of the measured cross section are found using the $\alpha$-OMPs of Ref.~\cite{McFadden66} (McFadden/Satchler) or a modified form of the $\alpha$-OMP from Ref.~\cite{Sauerwein11} (Sauerwein/Rauscher). Ultimately, the most suitable set of data to test the different $\alpha$-OMPs would be ($\alpha$,$\gamma$) cross sections at energies below the neutron emission threshold of \unit[$\approx 12$]{MeV}. 

\subsection{Astrophysical implications}

The astrophysical reaction rate is a central quantity in the $\gamma$-process and its theoretical description. However, the laboratory cross section, as measured in the present case, can only be used to directly calculate the stellar reaction rate, if the ground-state contribution $X~=~1$ \cite{Rauscher12}. For the present $^{168}$Yb($\alpha$,$\gamma$) reaction, the ground-state contribution is $X~=~0.16$ to $X~=~0.10$ for temperatures of \unit[2]{GK} to \unit[3]{GK} \cite{Nonsmoker}. In this case, experimentally determined laboratory cross sections might be used to constrain the nuclear-physics input used for stellar reaction rate predictions. 

The prediction of Ref.~\cite{Rauscher06} of $^{172}$Hf being a branching point in the $\gamma$-process path at a temperature of \unit[2]{GK} was obtained using the Non-Smoker default nuclear-physics input \cite{Nonsmoker}. In Sec.~\ref{sec:theory}, the experimental S factors are compared to this model prediction. It was found, that this calculation yields a fairly well description of the experimental data ($\chi^2=3.06$). Hence, one might conclude, that the present experiment confirms the model prediction of Ref.~\cite{Rauscher06} and $^{172}$Hf is a branching point in the $\gamma$-process path. Definite conclusions, however, are difficult to be drawn due to the complicated sensitivity of the cross section in the measured energy range to the $\gamma$ width and neutron width.

\subsection{Determination of $I_\gamma$ normalization factor for $^{171}$Lu}
\label{sec:intensity}
The absolute intensity of $\gamma$ rays emitted after the electron capture decay of $^{171}$Hf is unknown. However, the knowledge of decay parameters of the unstable reaction products is crucial, when performing activation experiments. Within the scope of this work it was possible to derive a normalization factor $N$ to determine the absolute $\gamma$-ray intensity $I_\gamma$ from the known relative intensities $I_{\mathrm{rel}}$ of three transitions for $^{171}$Lu. This was accomplished using the number of produced $^{171}$Hf nuclei at the end of the activation period, which could be derived during the data analysis of the $^{168}$Yb($\alpha$,n)$^{171}$Hf cross section determination, see Sec.~\ref{sec:dataanalysis}. Usually, absolute $\gamma$-ray intensities are given in literature as a product $N \times I_\mathrm{rel}(E_\gamma)$, where $N$ denotes a normalization factor and $I_\mathrm{rel}(E_\gamma)$ the relative intensity for a $\gamma$ ray emitted with energy $E_\gamma$. These relative intensities are usually normalized to one of the strongest $\gamma$ rays. This normalization factor $N$ was determined by

\begin{equation}
N = \frac{Y(E_\gamma)}{I_{\mathrm{rel}}(E_\gamma) \, \varepsilon(E_\gamma) \, \tau \, e^{-\lambda_\mathrm{Hf} \, \Delta t_\mathrm{wait}} \,  \left( 1 - e^{-\lambda_\mathrm{Hf}\,\Delta t_\mathrm{meas}} \right) N_{\mathrm{act}}^{\mathrm{Hf}}} \, .
\end{equation}

The relative $\gamma$-ray intensities of the $\gamma$ rays emitted subsequent to the electron capture decay of $^{171}$Hf normalized to \unit[$E_\gamma = 469.3$]{keV} are taken from Ref.~\cite{NNDC}, see Table~\ref{tab:intensity}. For two different $\alpha$-particle energies three $\gamma$-ray transitions following the electron capture decay of $^{171}$Hf, which decays with a half-life of \unit[(12.1 $\pm$ 0.4)]{h} \cite{NNDC} to $^{171}$Lu, could be used to derive the normalization factor $N$. Thus, in total six normalization factors could be derived, see Table~\ref{tab:intensity}. Subsequently, the weighted average of these factors was calculated. In a first step of this calculation, three average normalization factors were obtained by averaging over each $\gamma$-ray energy of the two different irradiated targets. In the end, the weighted average of these three normalization factors was calculated, which was found to be $N$~=~0.036~$\pm$~0.005. For each averaging step, only the independent errors were used for the weighting process. All systematic uncertainties, such as the uncertainty of the areal particle density, the number of impinging $\alpha$-particles, and the systematic uncertainty of the efficiency calibration was added afterwards by means of Gaussian error propagation. For each transition in $^{171}$Lu, this factor $N$ can be used to obtain the absolute $\gamma$-ray intensity per 100 decays of $^{171}$Hf, provided that the relative intensity is known. Table~\ref{tab:intensity} gives an overview of the results. 

\begin{table}[tb]
\centering
\caption{Measured normalization factors $N$ derived using the cross sections of the $^{168}$Yb($\alpha$,n)$^{171}$Hf reaction. These factors were determined at two different $\alpha$-particle energies for three different $\gamma$-ray energies each. $E_\gamma$ and $I_{\mathrm{rel}}$ taken from \cite{NNDC}. }
\begin{tabular}{c c c c c}
$E_\alpha$ [MeV]	&	$E_\gamma$ [keV]	&$I_{\mathrm{rel}}$&	$N$	&	$I_\gamma$ [\%] \\ \hline
\vspace{-0.3cm}
& & & &\\

13.76			&	347.2			&150 $\pm$ 20	&	0.035 $\pm$ 0.006	& 5.3 $\pm$ 1.2\\
			&	469.3			&100 $\pm$ 10	&	0.036 $\pm$ 0.006	& 3.6 $\pm$ 0.7\\
			&	1071.8			&148 $\pm$ 15	&	0.036 $\pm$ 0.006	& 5.3 $\pm$ 1.0\\
14.55			&	347.2			&150 $\pm$ 20	&	0.037 $\pm$ 0.006	& 5.5 $\pm$ 1.2\\
			&	469.3			&100 $\pm$ 10	&	0.034 $\pm$ 0.005	& 3.4 $\pm$ 0.6 \\
			&	1071.8			&148 $\pm$ 15 	&	0.037 $\pm$ 0.005	& 5.5 $\pm$ 0.9 \\ \hline
\end{tabular}
\label{tab:intensity}
\end{table}

Using the singles spectra emitted from the target irradiated with $\alpha$-particle energy of \unit[$E_\alpha~=~14.55$]{MeV}, nine relative intensities given in Ref.~\cite{NNDC} could be experimentally confirmed. For these cases, the absolute $\gamma$-ray intensities $I_\gamma$ were derived. For the calculation of these $I_\gamma$ values the averaged normalization factor was used. The results are shown in Table~\ref{tab:abs_intensity}.

\begin{table}[tb]
\centering
\caption{Absolute intensities $I_\gamma$ obtained for nine $\gamma$-ray transitions in $^{171}$Lu. The average normalization factor $N$ was used to calculate these from the relative intensities $I_{\mathrm{rel}}$ given in Ref.~\cite{NNDC}.}
\begin{tabular}{c c c}
$E_\gamma$ [keV]	&	$I_{\mathrm{rel}}$	&	$I_\gamma$ [\%] \\ \hline
\vspace{-0.3cm}
& &\\
113.1 	                &	20 $\pm$ 5              &	0.71 $\pm$ 0.19 \\
269.1   		&       40 $\pm$ 4              &	1.43 $\pm$ 0.19 \\
295.6           	&       137 $\pm$ 30            &	4.9 $\pm$ 1.2 \\
347.2           	&       150 $\pm$ 20            &	5.4 $\pm$ 0.9 \\
469.3           	&       100 $\pm$ 10            &	3.6 $\pm$ 0.5 \\
540.3                   &       35 $\pm$ 4       	&	1.3 $\pm$ 0.2 \\
662.2                   &       266 $\pm$ 30     	&  9.5 $\pm$ 1.4 \\
1071.8                  &       148 $\pm$ 15		&	5.3 $\pm$ 0.7 \\
1162.2                  &       33 $\pm$ 4       	&	1.2 $\pm$ 0.2 \\ \hline
\end{tabular}
\label{tab:abs_intensity}
\end{table}

\section{Summary}
Cross sections of the $^{168}$Yb($\alpha$,n)$^{171}$Hf and $^{168}$Yb($\alpha$,$\gamma$)$^{172}$Hf reactions were measured at energies of \unit[$E_0~=~12.9$]{MeV} to \unit[15.1]{MeV} using the activation technique. Two clover-type HPGe detectors were used to measure the ($\alpha$,n) cross section, whereas a LEPS detector was used for the determination of the ($\alpha$,$\gamma$) cross section. Additionally, the excellent agreement for one energy using the X-ray counting approach of Ref.~\cite{Kiss11} underlines once more the power of this method. 

Due to the complicated sensitivity of the cross section of both reactions to the $\alpha$ width, $\gamma$ width, and neutron width, it was not sufficient to vary the $\alpha$-OMP only.
It was found, that the best theoretical description for both reactions is given by the McFadden/Satchler $\alpha$-OMP of Ref.~\cite{McFadden66} combined with a standard Lorentzian approach for the photon strength function from Ref.~\cite{Brink57}. For this calculation the Generalized Superfluid Model of Ref.~\cite{Talys} was used for the nuclear level density and the Koning/Delaroche n-OMP of Ref.~\cite{Koning03}. A modification of the Sauerwein/Rauscher $\alpha$-OMP as used in Ref.~\cite{Rauscher12_1} yields a very good description of the experimental data as well. A comparison of the experimental data with (semi-)microscopic input parameters shows, that the absolute values are significantly underestimated, although the energy dependence is reproduced well. Since the default Non-Smoker calculation reproduces the experimental ($\alpha$,$\gamma$) data quite well, one might conclude, that the prediction of $^{172}$Hf being a branching point in the $\gamma$-process path is experimentally confirmed.

Finally, it was possible to derive the absolute $\gamma$-ray intensity for nine transitions following the electron capture decay of $^{171}$Hf, which were yet unknown. By using the cross section of the ($\alpha$,n) reaction, a normalization factor could be determined, which was subsequently used to derive the absolute $\gamma$-ray intensity from the relative intensities known from literature.

The results at hand underline once more the difficulties encountered in the determination of an $\alpha$-OMP, which is globally applicable. Further difficulties arise due to the various contributions of the particle widths and $\gamma$ width, that must be disentangled, when measuring at higher energies. The major uncertainties in astrophysical reaction rates involving $\alpha$-particles still arise from the $\alpha$-OMP. In order to achieve global improvements, more experimental data of $\alpha$-induced reactions at even lower $\alpha$-particle energies, \textit{i.e.} inside the Gamow window are desirable. Especially important are systematic measurements in order to achieve a more reliable description of the $\alpha$-OMP. In order to overcome the limitations of the activation technique, $e.\,g.$ the in-beam technique with HPGe detectors is promising, but very challenging for $\alpha$-induced reactions. Another promising approach to experimentally constrain the $\alpha$-OMP are experiments on elastic $\alpha$-scattering.

\section*{Acknowledgments}
The authors thank the accelerator staff and A. Heiske of PTB. We gratefully acknowledge H.-W. Becker and D. Rogalla of the Ruhr-Universität Bochum for their great assistance on RBS measurements. Furthermore, we thank P. Mohr for fruitful theoretical discussions, N. Warr for his helpful corrections, and A. Hennig and J. Winkens for their help on \textsc{Geant4} simulations. A.S. is supported by the Bonn-Cologne Graduate School of Physics and Astronomy. This project has been supported by the Deutsche Forschungsgemeinschaft under contracts ZI 510/5-1, INST 216/544-1, and the European Research Council starting Grant No. 203175, OTKA (PD104664). G.K. acknowledges support from the J\'anos Bolyai grant of the Hungarian Academy of Sciences.


\begin{thebibliography}{99}

\bibitem{B2FH}E.~M. Burbridge, G.~R. Burbridge, W.~A. Fowler, and F. Hoyle, Rev. Mod. Phys. 29 (1957) 547.

\bibitem{Kaeppeler11}F. K\"appeler, R. Gallino, S. Bisterzo, and W. Aoki, Rev. Mod. Phys. 83 (2011) 157.

\bibitem{Arnould07}M. Arnould, S. Goriely, and K. Takahashi, Phys. Rep. 450 (2007) 97.

\bibitem{Woosley78}S.~E. Woosley and W.~M. Howard, Astrophys. J. Suppl. 36 (1978) 285.

\bibitem{Arnould03}M. Arnould and S. Goriely, Phys. Rep. 384 (2003) 1.

\bibitem{Rayet95}M. Rayet, M. Hashimoti, N. Prantzos, and K. Nomoto, Astron. Astrophys. 298 (1995) 517.

\bibitem{Schatz98}H. Schatz, A. Aprahamian, J. G\"orres, M. Wiescher, T. Rauscher, J.~F. Rembges, F.-K. Thielemann, B. Pfeiffer, P. M\"oller, K.-L. Kratz, H. Herndl, B.~A. Brown, and H. Rebel, Phys. Rep. 294 (1998) 167.

\bibitem{Froehlich06}C. Fr\"ohlich, G. Mart\'{i}nez-Pinedo, M. Liebend\"orfer, F.-K. Thielemann, E. Bravo, W.~R. Hix, K. Langanke, and N.~T. Zinner, Phys. Rev. Lett. 96 (2006) 142502.

\bibitem{Goriely02}S. Goriely, J. Jos\'{e}, M. Hernanz, M. Rayet, and M. Arnould, Astron. Astrophys. 383 (2002) L27.

\bibitem{Arnould76}M. Arnould, Astron. Astrophys. 46 (1976) 117.

\bibitem{Travaglio11}C. Travaglio, F.~K. R\"opke, R. Gallino, and W. Hillebrandt, Astrophys. J. 739 (2011) 93.

\bibitem{Farouqi09}K. Farouqi, K.-L. Kratz, and B. Pfeiffer, Publ. Astron. Soc. Aust. 26 (2009) 194.

\bibitem{Hauser52}W. Hauser and H. Feshbach, Phys. Rev. 87 (1952) 366.

\bibitem{Yalcin09}C. Yal\c{c}in, N. \"Ozkan, S. Kutlu, Gy. Gy\"urky, J. Farkas, G.~G. Kiss, Zs. F\"ul\"op, A. Simon, E. Somorjai, and T. Rauscher, Phys. Rev. C 79 (2009) 065801.

\bibitem{Gyuerky10}Gy. Gy\"urky, Z. Elekes, J. Farkas, Zs. F\"ul\"op, Z. Hal\'{a}sz, G.~G. Kiss, E. Somorjai, T. Sz\"ucs, R.~T. G\"uray, N. \"Ozkan, C. Yal\c{c}in, and T. Rauscher, J. Phys. G 37 (2010) 115201.

\bibitem{Kiss11}G.~G. Kiss, T. Rauscher, T. Sz\"ucs, Zs. Kert\'{e}sz, Zs. F\"ul\"op, Gy. Gy\"urky, C. Fr\"ohlich, J. Farkas, Z. Elekes, and E. Somorjai, Phys. Lett. B 695 (2011) 419.

\bibitem{Filipescu11}D. Filipescu, V. Avrigeanu, T. Glodariu, C. Mihai, D. Bucurescu, M. Iva\c{s}cu, I. C\u{a}ta-Danil, L. Stroe, O. Sima, G. C\u{a}ta-Danil, D. Deleanu, D.~G. Ghi\c{t}\u{a}. N. M\u{a}rginean, R. M\u{a}rginean, A. Negret, S. Pascu, T. Sava, G. Suliman, and N.~V. Zamfir, Phys. Rev. C 83 (2011) 064609.

\bibitem{Sauerwein11}A. Sauerwein, H.-W. Becker, H. Dombrowski, M. Elvers, J. Endres, U. Giesen, J. Hasper, A. Hennig, L. Netterdon, T. Rauscher, D. Rogalla, K.~O. Zell, and A. Zilges, Phys. Rev. C 84 (2011) 045808.

\bibitem{Dillmann11}I. Dillmann, L. Coquard, C. Domingo-Pardo, F. K\"appeler, J. Marganiec, E. Uberseder, U. Giesen, A. Heiske, G. Feinberg, D. Hentschel, S. Hilpp, H. Leiste, T. Rauscher. F.-K. Thielemann, Phys. Rev. C 84 (2011) 015802.

\bibitem{Halasz12}Z. Hal\'{a}sz, Gy. Gy\"urky, J. Farkas, Zs. F\"ul\"op, T. Sz\"ucs, E. Somorjai, and T. Rauscher, Phys. Rev. C 85 (2012) 025804.

\bibitem{Harissopulos01}S. Harissopulos, E. Skreti, P. Tsagari, G. Souliotis, P. Demetriou, T. Paradellis, J.~W. Hammer, R. Kunz, C. Angulo, S. Goriely, and T. Rauscher, Phys. Rev. C 64 (2001) 055804.

\bibitem{Galanopoulos03}S. Galanopoulos, P. Demetriou, M. Kokkoris, S. Harissopulos, R. Kunz, M. Fey, J.~W. Hammer, Gy. Gy\"urky, Zs. F\"ul\"op, E. Somorjai, and S. Goriely, Phys. Rev. C 67 (2003) 015801.

\bibitem{Sauerwein12}A. Sauerwein, J. Endres, L. Netterdon, A. Zilges, V. Foteinou, G. Provatas, T. Konstantinopoulos, M. Axiotis, S.~F. Ashley, S. Harissopulos, and T. Rauscher, Phys. Rev. C 86 (2012) 035802.

\bibitem{Tsagari04}P. Tsagari, M. Kokkoris, E. Skreti, A.~G. Karydas, S. Harissopulos, T. Paradellis, and P. Demetriou, Phys. Rev. C 70 (2004) 015802.

\bibitem{Spyrou07}A. Spyrou, H.-W. Becker, A. Lagoyannis, S. Harissopulos, and C. Rolfs, Phys. Rev. C 76 (2007) 015802.

\bibitem{Rauscher06}T. Rauscher, Phys. Rev. C 73 (2006) 015804.

\bibitem{Rapp06}W. Rapp, J. G\"orres, M. Wiescher, H. Schatz, and F. K\"appeler, Astrophys. J. 653 (2006) 474.

\bibitem{Rauscher10}T. Rauscher, Phys. Rev. C, 81 (2010) 045807.

\bibitem{Rauscher12}T. Rauscher, Astrophys. J. Suppl. 201 (2012) 26.

\bibitem{PTB}H.~J. Brede, M. Cosack, G. Dietze, H. Gumpert, S. Guldbakke, R. Jahr, M. Kutscha, D. Schlegel-Bickmann, and H. Sch\"olermann, Nucl. Instr. Meth. 169 (1980) 349.

\bibitem{QCalc}National Nuclear Data Center, QCalc, \url{http://www.nndc.bnl.gov/qcalc}, last access 2013/07/16

\bibitem{Geant4}N. Agostinelli \textsl{et al.}, Nucl. Instr. Meth. A 506 (2003) 250.

\bibitem{Boettger}R. B\"ottger, private communication.

\bibitem{NNDC}National Nuclear Data Center, ENSDF database, \url{http://www.nndc.bnl.gov/ensdf}, last access 2013/07/16.


\bibitem{Kiss12}G.~G. Kiss, T. Sz\"ucs, Zs. T\"or\"ok, Z. Korkulu, Gy. Gy\"urky, Z. Hal\'{a}sz, Zs. F\"ul\"op, E. Somorjai, and T. Rauscher, Phys. Rev. C 86 (2012) 035801.

\bibitem{Semkov90}T.~M. Semkow, G. Mehmood, P.~P. Parekh, and M. Virgil, Nucl. Instr. Meth. A 290 (1990) 437.


\bibitem{Kiss11_1}G.~G. Kiss, T. Sz\"ucs, Gy. Gy\"urky, Zs. F\"ul\"op, J. Farkas, Zs. Kert\'{e}sz, E. Somorjai, M. Laubenstein, C. Fr\"ohlich, and T. Rauscher, Nucl. Phys. A 867 (2011) 52.

\bibitem{Lise}O. Tarasov and D. Bazin, LISE code version 9.3, available at \url{http://lise.nscl.msu.edu/lise.html}.

\bibitem{Rolfs88}C.~E. Rolfs and W.~S. Rodney, Cauldrons in the Cosmos, The University of Chicago Press, Chicago, 1988.

\bibitem{Rolfs06}C.~E. Rolfs, Nucl. Phys. News 16(2) (2006) 9.

\bibitem{Kettner06}K.~U. Kettner, H.-W. Becker, F. Strieder, and C. Rolfs, J. Phys. G:Nucl. Part. Phys. 32 (2006) 489.

\bibitem{Talys}A.~J. Koning, S. Hilaire, and M.~C. Duijvestijn, in \textsl{Proceedings of the International Conference on Nuclear Data for Science and Technology, April 22-27, 2007, Nice, France}, editors O. Bersillon, F. Gunsing, E. Bauge, R. Jacqmin, and S. Leray, EDP Sciences, 2008, p. 211-214; TALYS version 1.4 from www.talys.eu.

\bibitem{Watanabe58}S. Watanabe, Nucl. Phys. 8 (1958) 484.

\bibitem{McFadden66}L. McFadden and G.~R. Satchler, Nucl. Phys. 84 (1965) 177.

\bibitem{Demetriou02}P. Demetriou, C. Grama, and S. Goriely, Nucl. Phys. A 707 (2002) 253.

\bibitem{Brink57}D.~M. Brink, Nucl. Phys. 4 (1957) 215; P. Axel, Phys. Rev. 126 (1962) 671.

\bibitem{Kopecky90}J. Kopecky and M. Uhl, Phys. Rev. C 42 (1990) 1941.

\bibitem{Goriely02_1}S. Goriely and E. Khan, Nucl. Phys. A 706 (2002) 217.

\bibitem{Goriely04}S. Goriely, E. Khan, and M. Samyn, Nucl. Phys. A 739 (2004) 331.

\bibitem{Demetriou01}P. Demetriou and S. Goriely, Nucl. Phys. A 695 (2001) 95.

\bibitem{Goriely08}S. Goriely, S. Hilaire, and A.~J. Koning, Phys. Rev. C 78 (2008) 064307.

\bibitem{Koning08}A.~J. Koning, S. Hilaire, and S. Goriely, Nucl. Phys. A 810 (2008) 13.

\bibitem{Koning03}A.~J. Koning and J.~P. Delaroche, Nucl. Phys. A 713 (2003) 231.

\bibitem{Bauge01}E. Bauge, J.~P. Delaroche, and M. Girod, Phys. Rev. C 63 (2001) 024607.

\bibitem{Goriely07}S. Goriely and J.~P. Delaroche, Phys. Lett. B 653 (2007) 178183.

\bibitem{Capote09}R. Capote, M. Herman, P. Oblozinsky, P.~G. Young, S. Goriely, T. Belgya, A.~V. Ignatyuk, A.~J. Koning, S. Hilaire, V. Plujko, M. Avrigeanu, O. Bersillon, M.~B. Chadwick, T. Fukahori, S. Kailas, J. Kopecky, V.~M. Maslov, G. Reffo, M. Sin, E. Soukhovitskii, P. Talou, H. Yinlu, and G. Zhigang, Nucl. Data Sheets 110 (2009) 3107.

\bibitem{Nonsmoker}T. Rauscher and F.-K. Thielemann, At. Data Nucl. Data Tables 75 (2000) 1; 79 (2001) 47, accessed via \url{http://www.nucastro.org}.

\bibitem{Rauscher97}T. Rauscher, F.-K. Thielemann, and K.-L. Kratz, Phys. Rev. C 56 (1997) 1613.

\bibitem{Jeukenne77}J. Jeukenne, A. Lejeune, and C. Mahaux, Phys. Rev. C 16 (1977) 80

\bibitem{Kobos84}A.~M. Kobos, B.~A. Brown, R. Lindsay, and G.~R. Satchler, Nucl. Phys. A 425 (1984) 205.

\bibitem{Rauscher12_1}T. Rauscher, G.~G. Kiss, T. Sz\"ucs, Zs. F\"ul\"op, C. Fr\"ohlich, Gy. Gy\"urky, Z. Hal\'{a}sz, Zs. Kert\'{e}sz, and E. Somorjai, Phys. Rev. C 86 (2012) 015804.


\end{thebibliography}
\end{document}